\input harvmac
%
\message{S-Tables Macro v1.0, ACS, TAMU (RANHELP@VENUS.TAMU.EDU)}
%
%
\newhelp\stablestylehelp{You must choose a style between 0 and 3.}%
\newhelp\stablelinehelp{You should not use special hrules when
stretching
a table.}%
\newhelp\stablesmultiplehelp{You have tried to place an S-Table
inside another S-Table.  I would recommend not going on.}%
%
%
\newdimen\stablesthinline
\stablesthinline=0.4pt
\newdimen\stablesthickline
\stablesthickline=1pt
%
%
\newif\ifstablesborderthin
\stablesborderthinfalse
\newif\ifstablesinternalthin
\stablesinternalthintrue
\newif\ifstablesomit
\newif\ifstablemode
\newif\ifstablesright
\stablesrightfalse
%
%
\newdimen\stablesbaselineskip
\newdimen\stableslineskip
\newdimen\stableslineskiplimit
%
%
\newcount\stablesmode
\newcount\stableslines
\newcount\stablestemp
\stablestemp=3
\newcount\stablescount
\stablescount=0
\newcount\stableslinet
\stableslinet=0
%
%
%
\newcount\stablestyle
\stablestyle=0
%
%
\def\stablesleft{\quad\hfil}%
\def\stablesright{\hfil\quad}%
%
%
\catcode`\|=\active%
%
%
\newcount\stablestrutsize
\newbox\stablestrutbox
\setbox\stablestrutbox=\hbox{\vrule height10pt depth5pt width0pt}
\def\stablestrut{\relax\ifmmode%
                         \copy\stablestrutbox%
                       \else%
                         \unhcopy\stablestrutbox%
                       \fi}%
%
%
\newdimen\stablesborderwidth
\newdimen\stablesinternalwidth
\newdimen\stablesdummy
\newcount\stablesdummyc
\newif\ifstablesin
\stablesinfalse
%
%
%
%
%
\def\stablesadj{%
  \ifcase\stablestyle%
    \hbox to \hsize\bgroup\hss\vbox\bgroup%
  \or%
    \hbox to \hsize\bgroup\vbox\bgroup%
  \or%
    \hbox to \hsize\bgroup\hss\vbox\bgroup%
  \or%
    \hbox\bgroup\vbox\bgroup%
  \else%
    \errhelp=\stablestylehelp%
    \errmessage{Invalid style selected, using default}%
    \hbox to \hsize\bgroup\hss\vbox\bgroup%
  \fi}%
\def\stablesend{\egroup%
  \ifcase\stablestyle%
    \hss\egroup%
  \or%
    \hss\egroup%
  \or%
    \egroup%
  \or%
    \egroup%
  \else%
    \hss\egroup%
  \fi}%
\def\stablestart{%
  \ifstablesin%
    \errhelp=\stablesmultiplehelp%
    \errmessage{An S-Table cannot be placed within an S-Table!}%
  \fi
  \global\stablesintrue%
  \global\advance\stablescount by 1%
  \message{<S-Tables Generating Table \number\stablescount}%
  \begingroup%
  \stablestrutsize=\ht\stablestrutbox%
  \advance\stablestrutsize by \dp\stablestrutbox%
  \ifstablesborderthin%
    \stablesborderwidth=\stablesthinline%
  \else%
    \stablesborderwidth=\stablesthickline%
  \fi%
  \ifstablesinternalthin%
    \stablesinternalwidth=\stablesthinline%
  \else%
    \stablesinternalwidth=\stablesthickline%
  \fi%
  \tabskip=0pt%
  \stablesbaselineskip=\baselineskip%
  \stableslineskip=\lineskip%
  \stableslineskiplimit=\lineskiplimit%
  \offinterlineskip%
  \def\borderrule{\vrule width \stablesborderwidth}%
  \def\internalrule{\vrule width \stablesinternalwidth}%
  \def\thinline{\noalign{\hrule height \stablesthinline}}%
  \def\thickline{\noalign{\hrule height \stablesthickline}}%
  \def\trule{\omit\leaders\hrule height \stablesthinline\hfill}%
  \def\ttrule{\omit\leaders\hrule height \stablesthickline\hfill}%
  \def\tttrule##1{\omit\leaders\hrule height ##1\hfill}%
  \def\stablesel{&\omit\global\stablesmode=0%
    \global\advance\stableslines by 1\borderrule\hfil\cr}%
  \def\el{\stablesel&}%
  \def\elt{\stablesel\thinline&}%
  \def\eltt{\stablesel\thickline&}%
  \def\elttt##1{\stablesel\noalign{\hrule height ##1}&}%
  \def\elspec{&\omit\hfil\borderrule\cr\omit\borderrule&%
              \ifstablemode%
              \else%
                \errhelp=\stablelinehelp%
                \errmessage{Special ruling will not display properly}%
              \fi}%
  \def\stmultispan##1{\mscount=##1 \loop\ifnum\mscount>3
\stspan\repeat}%
  \def\stspan{\span\omit \advance\mscount by -1}%
  \def\multicolumn##1{\omit\multiply\stablestemp by ##1%
     \stmultispan{\stablestemp}%
     \advance\stablesmode by ##1%
     \advance\stablesmode by -1%
     \stablestemp=3}%
  \def\multirow##1{\stablesdummyc=##1\parindent=0pt\setbox0\hbox\bgroup%
    \aftergroup\emultirow\let\temp=}
  \def\emultirow{\setbox1\vbox to\stablesdummyc\stablestrutsize%
    {\hsize\wd0\vfil\box0\vfil}%
    \ht1=\ht\stablestrutbox%
    \dp1=\dp\stablestrutbox%
    \box1}%
%
  \def\stpar##1{\vtop\bgroup\hsize ##1%
     \baselineskip=\stablesbaselineskip%
     \lineskip=\stableslineskip%

\lineskiplimit=\stableslineskiplimit\bgroup\aftergroup\estpar\let\temp=}%
  \def\estpar{\vskip 6pt\egroup}%
  \def\stparrow##1##2{\stablesdummy=##2%
     \setbox0=\vtop to ##1\stablestrutsize\bgroup%
     \hsize\stablesdummy%
     \baselineskip=\stablesbaselineskip%
     \lineskip=\stableslineskip%
     \lineskiplimit=\stableslineskiplimit%
     \bgroup\vfil\aftergroup\estparrow%
     \let\temp=}%
  \def\estparrow{\vfil\egroup%
     \ht0=\ht\stablestrutbox%
     \dp0=\dp\stablestrutbox%
     \wd0=\stablesdummy%
     \box0}%
  \def|{\global\advance\stablesmode by 1&&&}%
  \def\|{\global\advance\stablesmode by 1&\omit\vrule width 0pt%
         \hfil&&}%
\def\vt{\global\advance\stablesmode
by 1&\omit\vrule width \stablesthinline%
          \hfil&&}%
  \def\vtt{\global\advance\stablesmode by 1&\omit\vrule width
\stablesthickline%
          \hfil&&}%
  \def\vttt##1{\global\advance\stablesmode by 1&\omit\vrule width ##1%
          \hfil&&}%
  \def\vtr{\global\advance\stablesmode by 1&\omit\hfil\vrule width%
           \stablesthinline&&}%
  \def\vttr{\global\advance\stablesmode by 1&\omit\hfil\vrule width%
            \stablesthickline&&}%
\def\vtttr##1{\global\advance\stablesmode
 by 1&\omit\hfil\vrule width ##1&&}%
  \stableslines=0%
  \stablesomitfalse}
\def\stablesdef{\bgroup\stablestrut\borderrule##\tabskip=0pt plus 1fil%
  &\stablesleft##\stablesright%
  &##\ifstablesright\hfill\fi\internalrule\ifstablesright\else\hfill\fi%
  \tabskip 0pt&&##\hfil\tabskip=0pt plus 1fil%
  &\stablesleft##\stablesright%
  &##\ifstablesright\hfill\fi\internalrule\ifstablesright\else\hfill\fi%
  \tabskip=0pt\cr%
  \ifstablesborderthin%
    \thinline%
  \else%
    \thickline%
  \fi&%
}%
\def\endtable{\advance\stableslines by 1\advance\stablesmode by 1%
   \message{- Rows: \number\stableslines, Columns:
\number\stablesmode>}%
   \stablesel%
   \ifstablesborderthin%
     \thinline%
   \else%
     \thickline%
   \fi%
   \egroup\stablesend%
\endgroup%
\global\stablesinfalse}
%

\overfullrule=0pt \abovedisplayskip=12pt plus 3pt minus 3pt
\belowdisplayskip=12pt plus 3pt minus 3pt

\noblackbox
\input epsf
\newcount\figno
\figno=0
\def\fig#1#2#3{
\par\begingroup\parindent=0pt\leftskip=1cm\rightskip=1cm\parindent=0pt
\baselineskip=11pt \global\advance\figno by 1 \midinsert
\epsfxsize=#3 \centerline{\epsfbox{#2}} \vskip 12pt
\centerline{{\bf Figure \the\figno:} #1}\par
\endinsert\endgroup\par}
\def\figlabel#1{\xdef#1{\the\figno}}

\def\np#1#2#3{Nucl. Phys. {\bf B#1} (#2) #3}

\def\IR{\relax{\rm I\kern-.18em R}}


\font\cmss=cmss10 \font\cmsss=cmss10 at 7pt
\def\rlx{\relax\leavevmode}
\def\inbar{\vrule height1.5ex width.4pt depth0pt}
\def\IC{\relax\,\hbox{$\inbar\kern-.3em{\rm C}$}}
\def\IN{\relax{\rm I\kern-.18em N}}
\def\IP{\relax{\rm I\kern-.18em P}}
\def\ZZ{\rlx\leavevmode\ifmmode\mathchoice{\hbox{\cmss Z\kern-.4em Z}}
 {\hbox{\cmss Z\kern-.4em Z}}{\lower.9pt\hbox{\cmsss Z\kern-.36em Z}}
 {\lower1.2pt\hbox{\cmsss Z\kern-.36em Z}}\else{\cmss Z\kern-.4em
 Z}\fi}
\def\IZ{\relax\ifmmode\mathchoice
{\hbox{\cmss Z\kern-.4em Z}}{\hbox{\cmss Z\kern-.4em Z}}
{\lower.9pt\hbox{\cmsss Z\kern-.4em Z}} {\lower1.2pt\hbox{\cmsss
Z\kern-.4em Z}}\else{\cmss Z\kern-.4em Z}\fi}

\def\narrowplus{\kern -.04truein + \kern -.03truein}
\def\narrowminus{- \kern -.04truein}
\def\narrowminussub{\kern -.02truein - \kern -.01truein}

\def\frac#1#2{{#1\over #2}}

\def\IZ{\relax\ifmmode\mathchoice
{\hbox{\cmss Z\kern-.4em Z}}{\hbox{\cmss Z\kern-.4em Z}}
{\lower.9pt\hbox{\cmsss Z\kern-.4em Z}} {\lower1.2pt\hbox{\cmsss
Z\kern-.4em Z}}\else{\cmss Z\kern-.4em Z}\fi}
\def\IB{\relax{\rm I\kern-.18em B}}
\def\IC{{\relax\hbox{$\inbar\kern-.3em{\rm C}$}}}
\def\ID{\relax{\rm I\kern-.18em D}}
\def\IE{\relax{\rm I\kern-.18em E}}
\def\IF{\relax{\rm I\kern-.18em F}}
\def\IG{\relax\hbox{$\inbar\kern-.3em{\rm G}$}}
\def\IGa{\relax\hbox{${\rm I}\kern-.18em\Gamma$}}
\def\IH{\relax{\rm I\kern-.18em H}}
\def\II{\relax{\rm I\kern-.18em I}}
\def\IK{\relax{\rm I\kern-.18em K}}
\def\IP{\relax{\rm I\kern-.18em P}}

\font\cmss=cmss10 \font\cmsss=cmss10 at 7pt
\def\IR{\relax{\rm I\kern-.18em R}}

\def\1{{\bf 1}}
\def\3{{\bf 3}}
\def\7{{\bf 7}}
\def\2{{\bf 2}}
\def\8{{\bf 8}}

%

%
%
\def\eqnn#1{\xdef #1{(\secsym\the\meqno)}\writedef{#1\leftbracket#1}%
\global\advance\meqno by1\wrlabeL#1}
\def\eqna#1{\xdef #1##1{\hbox{$(\secsym\the\meqno##1)$}}
\writedef{#1\numbersign1\leftbracket#1{\numbersign1}}%
\global\advance\meqno by1\wrlabeL{#1$\{\}$}}
\def\eqn#1#2{\xdef #1{(\secsym\the\meqno)}\writedef{#1\leftbracket#1}%
\global\advance\meqno by1$$#2\eqno#1\eqlabeL#1$$}



\lref\rBB{ K.~Becker and M.~Becker, ``${\cal M}$-Theory on
eight-manifolds,'', Nucl.\ Phys.\ {\bf B477} (1996) 155-167,
hep-th/9605053.}

\lref\DasguptaSS{ K.~Dasgupta, G.~Rajesh and S.~Sethi, ``M
theory, orientifolds and $G$-flux'', JHEP {\bf 9908} (1999) 023,
hep-th/9908088. }

\lref\beckerD{ K.~Becker and K.~Dasgupta, ``Heterotic strings
with torsion,'' JHEP {\bf 0211} (2002) 006, hep-th/0209077.}

\lref\rstrom{A.~Strominger, ``Superstrings with torsion'', Nucl.\
Phys.\ {\bf B274} (1986) 253-284.}

\lref\xenwit{M.~Dine, N.~Seiberg, X.~G.~Wen and E.~Witten,
``Nonperturbative effects on the string world sheet,'' Nucl.\
Phys.\ B {\bf 278} (1986) 769-789 ; ``Nonperturbative effects on
the string world sheet. 2,'' Nucl.\ Phys.\ B {\bf 289} (1987)
319-363. }

\lref\louisL{S.~Gurrieri, J.~Louis, A.~Micu and D.~Waldram,
``Mirror symmetry in generalized Calabi-Yau compactifications,''
Nucl.\ Phys.\ B {\bf 654}, 61 (2003), hep-th/0211102.}

\lref\HULL{C.~M.~Hull, ``Superstring compactifications with
torsion and space-time supersymmetry,'' {\it In Turin 1985,
Proceedings, Superunification and Extra Dimensions}, 347-375.}

\lref\hetcit{D.~J.~Gross, J.~A.~Harvey, E.~J.~Martinec and
R.~Rohm, ``The heterotic string,'' Phys.\ Rev.\ Lett.\ {\bf 54}
(1985) 502-505; ``Heterotic string theory. 1. The free heterotic
string,'' Nucl.\ Phys.\ B {\bf 256} (1985) 253-284; ``Heterotic
string theory. 2. The interacting heterotic string,'' Nucl.\
Phys.\ B {\bf 267} (1986) 75-124.}

\lref\sensethi{P.~S.~Aspinwall, ``$K3$ surfaces and string
duality,'', hep-th/9611137; A.~Sen and S.~Sethi, ``The mirror
transform of type I vacua in six dimensions,'' Nucl.\ Phys.\ B
{\bf 499} (1997) 45-54, hep-th/9703157; J.~de Boer, R.~Dijkgraaf,
K.~Hori, A.~Keurentjes, J.~Morgan, D.~R.~Morrison and S.~Sethi,
``Triples, fluxes, and strings,'' Adv.\ Theor.\ Math.\ Phys.\
{\bf 4} (2002) 995-1186, hep-th/0103170; D.~R.~Morrison and
S.~Sethi, ``Novel type I compactifications,'' JHEP {\bf 0201}
(2002) 032, hep-th/0109197.}

\lref\robbins{K.~Dasgupta, G.~Rajesh, D.~Robbins and S.~Sethi,
``Time-dependent warping, fluxes, and NCYM,'' JHEP {\bf 0303}
(2003) 041, hep-th/0302049; K.~Dasgupta and M.~Shmakova, ``On
branes and oriented B-fields,''
Nucl.\ Phys.\ B {\bf 675}, 205 (2003), hep-th/0306030.}

\lref\sewitten{N.~Seiberg, ``IR dynamics on branes and space-time
geometry,'' Phys.\ Lett.\ B {\bf 384} (1996) 81-85,
hep-th/9606017; N.~Seiberg and E.~Witten, ``Gauge dynamics and
compactification to three dimensions,'' hep-th/9607163.}

\lref\tseytlin{A.~A.~Tseytlin, ``On SO(32) heterotic - type I
superstring duality in ten dimensions,'' Phys.\ Lett.\ B {\bf
367} (1996) 84-90, hep-th/9510173; A.~A.~Tseytlin, ``Heterotic -
type I superstring duality and low-energy effective actions,''
Nucl.\ Phys.\ B {\bf 467} (1996) 383-398, hep-th/9512081.}


\lref\katzs{S.~Katz and E.~Sharpe, ``D-branes, open string vertex
operators, and Ext groups,'' Adv.\ Theor.\ Math.\ Phys.\  {\bf 6}
(2003) 979-1030, hep-th/0208104.}

\lref\rohmW{R.~Rohm and E.~Witten, ``The antisymmetric tensor
field in superstring theory,'' Annals Phys.\  {\bf 170} (1986)
454-489.}

\lref\bismut{J.-M. Bismut, ``A local index theorem for
non-K\"ahler manifolds,'' Math. \ Ann. {\bf 284} (1989) 681-699.}

\lref\hirze{F. Hirzebruch, {\it Topological Methods in Algebraic
Geometry}, Springer, Berlin, 1978.}

\lref\barsV{I.~Bars and M.~Visser, ``Number of massless fermion
families in superstring theory,'' Phys.\ Lett.\ B {\bf 163}
(1985) 118-122; ``Fermion families in superstring theory,''
USC-85/035.}

\lref\borcea{C. Borcea, ``$K3$ surfaces with involutions and
mirror pairs of Calabi-Yau manifolds,'' in {\it Mirror Manifolds
II}, ed. by B.~Greene and S.-T.~Yau.}

\lref\berkoozL{M.~Berkooz and R.~G.~Leigh, ``A D = 4 N = 1
orbifold of type I strings,'' Nucl.\ Phys.\ B {\bf 483} (1997)
187-208, hep-th/9605049.}

\lref\aspinwall{P.~S.~Aspinwall, ``Enhanced gauge symmetries and
$K3$ surfaces,'' Phys.\ Lett.\ B {\bf 357} (1995) 329-334,
hep-th/9507012.}

\lref\chs{A.~Strominger, ``Heterotic Solitons,'' Nucl.\ Phys.\ B
{\bf 343}, 167 (1990) [Erratum-ibid.\ B {\bf 353}, 565 (1991)];
C.~G.~Callan, J.~A.~Harvey and A.~Strominger, ``World Sheet
Approach To Heterotic Instantons And Solitons,'' Nucl.\ Phys.\ B
{\bf 359}, 611 (1991).}

\lref\adopt{J.~D.~Edelstein, K.~Oh and R.~Tatar,
``Orientifold, geometric transition and large N duality for SO/Sp gauge  theories,''
JHEP {\bf 0105}, 009 (2001), hep-th/0104037;
K.~Dasgupta, K.~Oh and R.~Tatar, {``Geometric
transition, large N dualities and MQCD dynamics,''} Nucl.\ Phys.\
B {\bf 610}, 331 (2001), hep-th/0105066; {``Open/closed string
dualities and Seiberg duality from geometric transitions in
M-theory,''} JHEP {\bf 0208}, 026 (2002), hep-th/0106040;
K.~Dasgupta, K.~h.~Oh, J.~Park and R.~Tatar, ``Geometric
transition versus cascading solution,'' JHEP {\bf 0201}, 031
(2002), hep-th/0110050;
K.~h.~Oh and R.~Tatar,
``Duality and confinement in N = 1 supersymmetric theories from geometric  transitions,''
Adv.\ Theor.\ Math.\ Phys.\  {\bf 6}, 141 (2003), hep-th/0112040.}

\lref\gillardetal{J.~Gillard, G.~Papadopoulos, and D.~Tsimpis,
``Anomaly, fluxes, and $(2,0)$ heterotic-string
compactifications,'' hep-th/0304126.}

\lref\spindel{Ph.~Spindel, A.~Sevrin, W.~Troost, and
A.~van~Proeyen, ``Complex structures on parallelised group
manifolds and supersymmetric sigma models,'' Phys. Lett. B {\bf
206} (1988) 71-74; ``Extended supersymmetric sigma models on
group manifolds: I,'' Nucl. Phys. B {\bf 308} (1988) 662-698;
``Extended supersymmetric sigma models on group manifolds: II,''
Nucl. Phys. B {\bf 311} (1988) 465-492.}

\lref\brin{V.~Brinzanescu and R.~Moraru, ``Holomorphic rank-2
vector bundles on non-K\"ahler elliptic surfaces,''
math.AG/0306191; ``Stable bundles on non-K\"ahler elliptic
surfaces,'' math.AG/0306192; ``Twisted Fourier-Mukai transforms
and bundles on non-K\"ahler elliptic surfaces,'' math.AG/0309031.}

\lref\ivanov{J.~Gutowski, S.~Ivanov, and G.~Papadopoulos,
``Deformations of generalized calibrations and compact
non-K\"ahler manifolds with vanishing first Chern class,''
math.DG/0205012.}

\lref\font{G.~Aldazabal, A.~Font, L.~E.~Ibanez and G.~Violero,
{``D = 4, N = 1, type IIB orientifolds,''} Nucl.\ Phys.\ B {\bf
536}, 29 (1998), hep-th/9804026.}

\lref\urangafont{G.~Aldazabal, A.~Font, L.~E.~Ibanez,
A.~M.~Uranga and G.~Violero, {``Non-perturbative heterotic D =
6,4, N = 1 orbifold vacua,''} Nucl.\ Phys.\ B {\bf 519}, 239
(1998), hep-th/9706158.}

\lref\nappiS{M.~Dine, V.~Kaplunovsky, M.~L.~Mangano, C.~Nappi and
N.~Seiberg, {``Superstring Model Building,''} Nucl.\ Phys.\ B
{\bf 259}, 549 (1985).}

\lref\glashow{H.~Georgi and S.~L.~Glashow, ``Unity Of All
Elementary Particle Forces,'' Phys.\ Rev.\ Lett.\  {\bf 32}, 438
(1974).}

\lref\slansky{R.~Slansky, ``Group Theory For Unified Model
Building,'' Phys.\ Rept.\  {\bf 79}, 1 (1981).}

\lref\curioLu{G.~Curio and A.~Krause, ``Enlarging the parameter
space of heterotic M-theory flux compactifications to
phenomenological viability,'' hep-th/0308202.}

\lref\dabhull{A.~Dabholkar and C.~Hull,
``Duality twists, orbifolds, and fluxes,''
JHEP {\bf 0309}, 054 (2003), hep-th/0210209.}

\lref\prokS{S.~Prokushkin, Private Communications.}

\lref\buchkov{E.~I.~Buchbinder and B.~A.~Ovrut,
``Vacuum stability in heterotic M-theory,'' hep-th/0310112.}

\lref\warner{M.~Gunaydin and N.~P.~Warner, ``The G2 invariant
compactifications in eleven-dimensional supergravity,'' Nucl.\
Phys.\ B {\bf 248}, 685 (1984); P.~van Nieuwenhuizen and
N.~P.~Warner, ``New compactifications of ten-dimensional and
eleven-dimensional supergravity on manifolds which are not direct
products,'' Commun.\ Math.\ Phys.\  {\bf 99}, 141 (1985).}

\lref\nemes{D.~Nemeschansky and S.~Yankielowicz, ``Critical d
imension of string theories in curved space,'' Phys.\ Rev.\
Lett.\  {\bf 54}, 620 (1985) [Erratum-ibid.\  {\bf 54}, 1736
(1985)]; D.~Friedan, Z.~a.~Qiu and S.~H.~Shenker,
``Superconformal invariance in two dimensions and the tricritical
Ising model,'' Phys.\ Lett.\ B {\bf 151}, 37 (1985).}

\lref\bbdgs{K.~Becker, M.~Becker, P.~S.~Green, K.~Dasgupta and
E.~Sharpe, ``Compactifications of heterotic strings on
non-K\"ahler complex manifolds. II,'' Nucl. \ Phys. \ B {\bf 678}, 19 (2004),
 hep-th/0310058.}

\lref\senn{D.~Nemeschansky and A.~Sen, ``Conformal invariance of
supersymmetric sigma models on Calabi-Yau manifolds,'' Phys.\
Lett.\ B {\bf 178}, 365 (1986); P.~Candelas, M.~D.~Freeman,
C.~N.~Pope, M.~F.~Sohnius and K.~S.~Stelle, ``Higher order
corrections to supersymmetry and compactifications of the
heterotic string,'' Phys.\ Lett.\ B {\bf 177}, 341 (1986).}

\lref\rDJM{K. Dasgupta, D. P. Jatkar and S. Mukhi,
``Gravitational couplings and $Z_2$ orientifolds'', Nucl. Phys.
{\bf B523} (1998) 465-484, hep-th/9707224; J.~F.~Morales,
C.~A.~Scrucca and M.~Serone, ``Anomalous couplings for D-branes
and O-planes,'' Nucl.\ Phys.\ B {\bf 552} (1999) 291-315,
hep-th/9812071; B.~J.~Stefanski, ``Gravitational couplings of
D-branes and O-planes,'' Nucl.\ Phys.\ B {\bf 548} (1999)
275-290, hep-th/9812088. }

\lref\nemanja{N.~Kaloper and R.~C.~Myers,  ``The O(dd) story of
massive supergravity,'' JHEP {\bf 9905} (1999) 010,
hep-th/9901045; G.~Curio, A.~Klemm, B.~Kors and D.~Lust, ``Fluxes
in heterotic and type II string compactifications,'' Nucl.\
Phys.\ B {\bf 620} (2002) 237-258, hep-th/0106155; J.~Louis and
A.~Micu, ``Heterotic string theory with background fluxes,''
Nucl.\ Phys.\ B {\bf 626} (2002) 26-52, hep-th/0110187.}

\lref\kapulov{V.~Kaplunovsky, J.~Louis and S.~Theisen, ``Aspects
of duality in N=2 string vacua,'' Phys.\ Lett.\ B {\bf 357}
(1995) 71-75, hep-th/9506110.}

\lref\nati{N.~Seiberg, ``Observations on the moduli space of
superconformal field theories,'' Nucl.\ Phys.\ B {\bf 303} (1988)
286-304; A.~Ceresole, R.~D'Auria, S.~Ferrara and A.~Van Proeyen,
``Duality transformations in supersymmetric Yang-Mills theories
coupled to supergravity,'' Nucl.\ Phys.\ B {\bf 444} (1995)
92-124, hep-th/9502072.}

\lref\narainJ{K.~S.~Narain, ``New heterotic string theories in
uncompactified dimensions $< 10$,'' Phys.\ Lett.\ B {\bf 169}
(1986) 41-46; K.~S.~Narain, M.~H.~Sarmadi and E.~Witten, ``A note
on toroidal compactification of heterotic string theory,'' Nucl.\
Phys.\ B {\bf 279} (1987) 369-379; J.~Maharana and J.~H.~Schwarz,
``Noncompact symmetries in string theory,'' Nucl.\ Phys.\ B {\bf
390} (1993) 3-32 (1993), hep-th/9207016.}

\lref\senM{A.~Sen, ``A note on enhanced gauge symmetries in M-
and string theory,'' JHEP {\bf 9709} (1997) 001, hep-th/9707123.}

\lref\rBUSH{T. Buscher, ``Quantum corrections and extended
supersymmetry in new sigma models'', Phys. Lett. {\bf B159}
(1985) 127-130; ``A symmetry of the string background field
equations,'' Phys. Lett. B {\bf 194} (1987) 59-62; ``Path
integral derivation of quantum duality in nonlinear sigma
models'', Phys. Lett. B {\bf 201} (1988) 466-472.}

\lref\rKKL{E. Kiritsis, C. Kounnas and D. Lust, ``A large class
of new gravitational and axionic backgrounds for four-dimensional
superstrings'', Int. J. Mod. Phys. {\bf A9} (1994) 1361-1394,
hep-th/9308124. }

\lref\papado{J.~Gutowski, S.~Ivanov
and G.~Papadopoulos, ``Deformations of generalized calibrations
and compact non-K\"ahler manifolds with vanishing first Chern
class,'' math.dg/0205012.}

\lref\gauntlett{J.~P.~Gauntlett, N.~w.~Kim, D.~Martelli and
D.~Waldram, ``Fivebranes wrapped on SLAG three-cycles and related
geometry,'' JHEP {\bf 0111} (2001) 018, hep-th/0110034.
J.~P.~Gauntlett, D.~Martelli, S.~Pakis and D.~Waldram,
``G-structures and wrapped NS5-branes,'' hep-th/0205050;
J.~P.~Gauntlett, D.~Martelli and D.~Waldram, ``Superstrings with
intrinsic torsion,'' hep-th/0302158.}

\lref\sengimon{A.~Sen, ``A non-perturbative description of the
Gimon-Polchinski orientifold,'' Nucl.\ Phys.\ B {\bf 489} (1997)
139-159, hep-th/9611186; A.~Sen, ``F-theory and the
Gimon-Polchinski orientifold,'' Nucl.\ Phys.\ B {\bf 498} (1997)
135-155, hep-th/9702061.}

\lref\gimpol{E.~G.~Gimon and J.~Polchinski, ``Consistency
conditions for orientifolds and D-manifolds,'' Phys.\ Rev.\ D
{\bf 54} (1996) 1667-1676, hep-th/9601038.}

\lref\kst{S. Giddings, S. Kachru and J. Polchinski, ``Hierarchies
from fluxes in string compactifications,'' hep-th/0105097;
S.~Kachru, M.~B.~Schulz and S.~Trivedi, ``Moduli stabilization
from fluxes in a simple IIB orientifold,'' hep-th/0201028;
A.~R.~Frey and J.~Polchinski, ``N = 3 warped compactifications,
'' Phys.\ Rev.\ {\bf D65} (2002) 126009, hep-th/0201029. }

\lref\pktspto{ S.~Gurrieri and A.~Micu, ``Type IIB theory on
half-flat manifolds,'' hep-th/0212278.}
\lref\pktsptt{P.~K.~Tripathy and S.~P.~Trivedi, Compactifications
with flux on $K3$ and tori,'' hep-th/0301139.}

\lref\gates{ S.~J.~Gates, ``Superspace formulation of new
nonlinear sigma models,'' Nucl.\ Phys.\ B {\bf 238} (1984)
349-366; S.~J.~Gates, C.~M.~Hull and M.~Rocek, ``Twisted
multiplets and new supersymmetric nonlinear sigma models,''
Nucl.\ Phys.\ B {\bf 248} (1984) 157-186; S.~J.~Gates, S.~Gukov
and E.~Witten, ``Two-dimensional supergravity theories from
Calabi-Yau four-folds,'' Nucl.\ Phys.\ B {\bf 584} (2000) 109-148,
hep-th/0005120.}

\lref\rkehagias{A. Kehagias, ``New type IIB vacua and their
F-theory interpretation,'' Phys. Lett. B {\bf 435} (1998) 337-342,
hep-th/9805131. }

\lref\GukovYA{ S.~Gukov, C.~Vafa and E.~Witten, ``CFT's from
Calabi-Yau four-folds,'' Nucl.\ Phys.\ B {\bf 584} (2000) 69-108,
erratum ibid {\bf 608} (2001) 477-478, hep-th/9906070. }
\lref\BeckerPM{ K.~Becker and M.~Becker, ``Supersymmetry
breaking, ${\cal M}$-theory and fluxes'', JHEP {\bf 0107} (2001)
038, hep-th/0107044. }
\lref\DineRZ{ M.~Dine, R.~Rohm, N.~Seiberg and E.~Witten,
``Gluino condensation in superstring models,'' Phys.\ Lett.\ B
{\bf 156} (1985) 55-60.}
\lref\FreyHF{ A.~R.~Frey and J.~Polchinski, ``N = 3 warped
compactifications,'' Phys.\ Rev.\ D {\bf 65} (2002) 126009,
hep-th/0201029.}
\lref\CurioAE{ G.~Curio, A.~Klemm, B.~Kors and D.~Lust, ``Fluxes
in heterotic and type II string compactifications,'' Nucl.\
Phys.\ B {\bf 620} (2002) 237-258, hep-th/0106155.}

\lref\Vafawitten{ C.~Vafa and E.~Witten, ``A one loop test of
string duality,'' Nucl.\ Phys.\ B {\bf 447} (1995) 261-270,
hep-th/9505053.}

\lref\SethiVW{ S.~Sethi, C.~Vafa and E.~Witten, ``Constraints on
low-dimensional string compactifications,'' Nucl.\ Phys.\ B {\bf
480} (1996) 213-224, hep-th/9606122.}

\lref\HananyK{ A.~Hanany and B.~Kol, ``On orientifolds, discrete
torsion, branes and M theory,'' JHEP {\bf 0006} (2000) 013,
hep-th/0003025.}

\lref\Ganor{ O.~J.~Ganor, ``Compactification of tensionless
string theories,'' hep-th/9607092.}

\lref\DMtwo{ K.~Dasgupta and S.~Mukhi,  ``A note on
low-dimensional string compactifications,'' Phys.\ Lett.\ B {\bf
398} (1997) 285-290, hep-th/9612188.}

\lref\ShiuG{ B.~R.~Greene, K.~Schalm and G.~Shiu,  ``Warped
compactifications in M and F theory,'' Nucl.\ Phys.\ B {\bf 584}
(2000) 480-508, hep-th/0004103.}

\lref\harmoni{ G.~W.~Gibbons and P.~J.~Ruback, ``The hidden
symmetries of multicenter metrics,'' Commun.\ Math.\ Phys.\ {\bf
115} (1988) 267-300; N.~S.~Manton and B.~J.~Schroers, ``Bundles
over moduli spaces and the quantization of BPS monopoles,''
Annals Phys.\  {\bf 225} (1993) 290-338; A.~Sen, ``Dyon -
monopole bound states, selfdual harmonic forms on the multi -
monopole moduli space, and SL(2,Z) invariance in string theory,''
Phys.\ Lett.\ B {\bf 329} (1994) 217-221, hep-th/9402032.}

\lref\mesO{P.~Meessen and T.~Ortin, ``An Sl(2,Z) multiplet of
nine-dimensional type II supergravity theories,'' Nucl.\ Phys.\ B
{\bf 541} (1999) 195-245, hep-th/9806120; E. Bergshoeff, C. Hull
and T. Ortin, ``Duality in the type II superstring effective
action,'' \np{451} {1995}{547-578}, hep-th/9504081; S.~F.~Hassan,
``T-duality, space-time spinors and R-R fields in curved
backgrounds,'' Nucl.\ Phys.\ B {\bf 568} (2000) 145-161,
hep-th/9907152.}

\lref\greenS{M.~B.~Green and J.~H.~Schwarz, ``Superstring
interactions,'' Nucl.\ Phys.\ B {\bf 218} (1983) 43-88.}

\lref\SmitD{ B.~de Wit, D.~J.~Smit and N.~D.~Hari Dass,
``Residual supersymmetry of compactified D = 10 supergravity,''
Nucl.\ Phys.\ B {\bf 283} (1987) 165-191.}

\lref\greene{B.~R.~Greene, A.~D.~Shapere, C.~Vafa and S.~T.~Yau,
``Stringy cosmic strings and noncompact Calabi-Yau manifolds,''
Nucl.\ Phys.\ B {\bf 337} (1990) 1-36.}

\lref\polci{J.~Polchinski, ``Tensors from $K3$ orientifolds,''
Phys.\ Rev.\ D {\bf 55} (1997) 6423-6428, hep-th/9606165.}

\lref\dabholkarP{A.~Dabholkar and J.~Park, ``An orientifold of
type-IIB theory on $K3$,'' Nucl.\ Phys.\ B {\bf 472} (1996)
207-220, hep-th/9602030; ``Strings on orientifolds,'' Nucl.\
Phys.\ B {\bf 477} (1996) 701-714, hep-th/9604178; ``A note on
orientifolds and F-theory,'' Phys.\ Lett.\ B {\bf 394} (1997)
302-306, hep-th/9607041.}

\lref\nikulin{V. Nikulin, ``Discrete reflection groups in
Lobachevsky spaces and algebraic surfaces,'' pp.~654-671 in {\it
Proceedings of the International Congress of Mathematicians},
Berkeley (1986) 654.}

\lref\samref{S.~Kachru and C.~Vafa, ``Exact results for N=2
compactifications of heterotic strings,'' Nucl.\ Phys.\ B {\bf
450} (1995) 69-89, hep-th/9505105; M.~Bershadsky,
K.~A.~Intriligator, S.~Kachru, D.~R.~Morrison, V.~Sadov and
C.~Vafa, ``Geometric singularities and enhanced gauge
symmetries,'' Nucl.\ Phys.\ B {\bf 481} (1996) 215-252,
hep-th/9605200; P.~S.~Aspinwall and M.~Gross,
``Heterotic-heterotic string duality and multiple $K3$
fibrations,'' Phys.\ Lett.\ B {\bf 382} (1996) 81-88,
hep-th/9602118.}

\lref\duff{M.~J.~Duff, R.~Minasian and E.~Witten, ``Evidence for
heterotic/heterotic duality,'' Nucl.\ Phys.\ B {\bf 465} (1996)
413-438, hep-th/9601036.}

\lref\morvaf{D.~R.~Morrison and C.~Vafa, ``Compactifications of
F-Theory on Calabi--Yau threefolds -- I,'' Nucl.\ Phys.\ B {\bf
473} (1996) 74-92, hep-th/9602114; ``Compactifications of
F-Theory on Calabi--Yau threefolds -- II,'' Nucl.\ Phys.\ B {\bf
476} (1996) 437-469, hep-th/9603161.}

\lref\dasF{ K.~Dasgupta and S.~Mukhi, ``F-theory at constant
coupling,'' Phys.\ Lett.\ B {\bf 385} (1996) 125-131,
hep-th/9606044.}

\lref\PapaDI{ S.~Ivanov and G.~Papadopoulos, ``A no-go theorem
for string warped compactifications,'' Phys.\ Lett. B {\bf 497}
(2001) 309-316, hep-th/0008232.}

\lref\DineSB{M.~Dine and N.~Seiberg, ``Couplings and scales in
superstring models,'' Phys.\ Rev.\ Lett.\  {\bf 55} (1985)
366-369.}

\lref\olgi{O. DeWolfe and S. B. Giddings, ``Scales and
hierarchies in warped compactifications and brane worlds,''
hep-th/0208123.}

\lref\hellermanJ{S.~Hellerman, J.~McGreevy and B.~Williams,
``Geometric constructions of non-geometric string theories,''
hep-th/0208174.}

\lref\WIP{K. Becker, M. Becker, K. Dasgupta, work in progress.}

\lref\tatar{K.~Dasgupta, K.~h.~Oh, J.~Park and R.~Tatar,
``Geometric transition versus cascading solution,'' JHEP {\bf
0201} (2002) 031, hep-th/0110050.}

\lref\wipro{Work in progress.}

\lref\civ{F.~Cachazo, B.~Fiol, K.~A.~Intriligator, S.~Katz and
C.~Vafa, ``A geometric unification of dualities,'' Nucl.\ Phys.\
B {\bf 628} (2002) 3-78, hep-th/0110028.}

\lref\renata{K.~Dasgupta, C.~Herdeiro, S.~Hirano and R.~Kallosh,
``D3/D7 inflationary model and M-theory,'' Phys.\ Rev.\ D {\bf
65} (2002) 126002, hep-th/0203019.}

\lref\carlos{R.~Kallosh, ``N = 2 supersymmetry and de Sitter
space,'' hep-th/0109168; C.~Herdeiro, S.~Hirano and R.~Kallosh,
``String theory and hybrid inflation / acceleration,'' JHEP {\bf
0112} (2001) 027, hep-th/0110271.}

\lref\kklt{S.~Kachru, R.~Kallosh, A.~Linde and S.~P.~Trivedi,
{``De Sitter vacua in string theory,''} Phys.\ Rev.\ D {\bf 68},
046005 (2003), hep-th/0301240; C.~P.~Burgess, R.~Kallosh and
F.~Quevedo, {``De Sitter string vacua from supersymmetric
D-terms,''}, hep-th/0309187.}

\lref\trivsham{S.~Kachru, R.~Kallosh, A.~ Linde, J.~Maldacena,
L.~ McAllister, S.~Trivedi, work in progress.}

\lref\polchinski{ J.~Polchinski, {\it String Theory. Vol. 2:
Superstring Theory And Beyond}.}

\lref\taylor{E.~Cremmer, S.~Ferrara, L.~Girardello and A.~Van
Proeyen, ``Yang-Mills theories with local supersymmetry:
lagrangian, transformation laws and superhiggs effect,'' Nucl.\
Phys.\ B {\bf 212} (1983) 413-442; T.~R.~Taylor and C.~Vafa, ``RR
Flux on Calabi-Yau and partial supersymmetry breaking,'' Phys.\
Lett.\ B {\bf 474} (2000) 130-137, hep-th/9912152.}

\lref\guko{S.~Gukov, C.~Vafa and E.~Witten, ``CFT's from
Calabi-Yau four-folds,'' Nucl.\ Phys.\ B {\bf 584} (2000) 69-108,
[Erratum-ibid.\ B {\bf 608} (2001) 477-478], hep-th/9906070.}

\lref\hullwitten{ C.~M.~Hull and E.~Witten, ``Supersymmetric
sigma models and the heterotic string,'' Phys.\ Lett.\ B {\bf
160} (1985) 398-402; A.~Sen, ``Local gauge and lorentz invariance
of the heterotic string theory,'' Phys.\ Lett.\ B {\bf 166}
(1986) 300-304; ``The heterotic string in arbitrary background
fields,'' Phys.\ Rev.\ D {\bf 32} (1985) 2102-2112; ``Equations
of motion for the heterotic string theory from the conformal
invariance of the sigma model,'' Phys.\ Rev.\ Lett.\ {\bf 55}
(1985) 1846-1849.}

\lref\dasmukhi{ K.~Dasgupta and S.~Mukhi, ``F-theory at constant
coupling,'' Phys.\ Lett.\ B {\bf 385} (1996) 125-131,
hep-th/9606044.}

\lref\vafasen{ C.~Vafa, ``Evidence for F-theory,'' Nucl.\ Phys.\
B {\bf 469} (1996) 403-418, hep-th/9602022; A.~Sen, ``F-theory
and orientifolds,'' Nucl.\ Phys.\ B {\bf 475} (1996) 562-578,
hep-th/9605150; T.~Banks, M.~R.~Douglas and N.~Seiberg, ``Probing
F-theory with branes,'' Phys.\ Lett.\ B {\bf 387} (1996) 278-281,
hep-th/9605199.}

\lref\zwee{ M.~R.~Gaberdiel and B.~Zwiebach, ``Exceptional groups
from open strings,'' Nucl.\ Phys.\ B {\bf 518} (1998) 151-172,
hep-th/9709013.}

\lref\howi{P.~Horava and E.~Witten, ``Heterotic and type I
dynamics from eleven dimensions,'' Nucl.\ Phys.\ B {\bf 460}
(1996) 506-524, hep-th/9510209.}

\lref\howii{P.~Horava and E.~Witten, ``Eleven-dimensional
supergravity on a manifold with boundary ,'' Nucl.\ Phys.\ B {\bf
475} (1996) 94-114, hep-th/9603142.}

\lref\mps{G.~Moore, G.~Peradze and N.~Saulina, ``Instabilities in
heterotic M theory induced by open membrane instantons ,'' Nucl.\
Phys.\ B {\bf 607} (2001) 117-154, hep-th/0012104.}

\lref\cukri{G.~Curio and A.~Krause, ``G fluxes and nonperturbative
stabilization of heterotic M theory ,'' Nucl.\ Phys.\ B {\bf 643}
(2002) 131-156, hep-th/0108220.}

\lref\cukrii{G.~Curio and A.~Krause, ``Enlarging the parameter
space of heterotic M theory flux compactifications to
phenomenological viability,'' hep-th/0308202.}

\lref\wittenstrong{E.~Witten, ``Strong coupling expansion of
Calabi-Yau compactification,'' Nucl.\ Phys.\ B {\bf 471} (1996)
135-158, hep-th/9602070.}

\lref\cukriii{G.~Curio and A.~Krause, ``Four flux and warped
heterotic M theory compactifications,'' Nucl.\ Phys.\ B {\bf 602}
(2001) 172-200, hep-th/0012152.}

\lref\bckr{M.~Becker, G.~Curio and A.~Krause,``Moduli
stabilization and De Sitter vacua from heterotic M-theory'', to
appear.}

\lref\callan{ C.~G.~Callan, E.~J.~Martinec, M.~J.~Perry and
D.~Friedan, ``Strings in background fields,'' Nucl.\ Phys.\ B
{\bf 262} (1985) 593-609; A.~Sen, ``Equations of motion for the
heterotic string theory from the conformal invariance of the
sigma model,'' Phys.\ Rev.\ Lett.\  {\bf 55} (1985) 1846-1849;
``The heterotic string in arbitrary background field,'' Phys.\
Rev.\ D {\bf 32} (1985) 2102-2112.}

\lref\WittenBS{ E.~Witten, ``Toroidal compactification without
vector structure,'' JHEP {\bf 9802} (1998) 006, hep-th/9712028.}

\lref\carluest{ G.~L.~Cardoso, G.~Curio, G.~Dall'Agata, D.~Lust,
P.~Manousselis and G.~Zoupanos, ``Non-K\"ahler string backgrounds
and their five torsion classes,''
Nucl.\ Phys.\ B {\bf 652}, 5 (2003),
hep-th/0211118.}

\lref\grifhar{ P.~ Griffiths and J.~ Harris {\it ``Principles of
Algebraic Geometry,''} Wiley Classics Library.}

\lref\hanany{A.~Hanany and A.~Zaffaroni, ``On the realization of
chiral four-dimensional gauge theories using  branes,'' JHEP {\bf
9805} (1998) 001, hep-th/9801134; A.~Hanany and A.~M.~Uranga,
``Brane boxes and branes on singularities,'' JHEP {\bf 9805}
(1998) 013, hep-th/9805139.}

\lref\dasmuk{B.~Andreas, G.~Curio and D.~Lust, ``The
Neveu-Schwarz five-brane and its dual geometries,'' JHEP {\bf
9810} (1998) 022, hep-th/9807008; K.~Dasgupta and S.~Mukhi,
``Brane constructions, conifolds and M-theory,'' Nucl.\ Phys.\ B
{\bf 551} (1999) 204-228, hep-th/9811139.}

\lref\zwigab{A.~Johansen, ``A comment on BPS states in F-theory
in 8 dimensions,'' Phys.\ Lett.\ B {\bf 395} (1997) 36-41,
hep-th/9608186; M.~R.~Gaberdiel and B.~Zwiebach, ``Exceptional
groups from open strings,'' Nucl.\ Phys.\ B {\bf 518} (1998)
151-172, hep-th/9709013; M.~R.~Gaberdiel, T.~Hauer and
B.~Zwiebach, ``Open string-string junction transitions,'' Nucl.\
Phys.\ B {\bf 525} (1998) 117-145, hep-th/9801205.}

\lref\gsw{M.~B.~Green, J.~H.~Schwarz and E.~Witten, {\it
Superstring Theory},  Vol. 1, 2; J.~Polchinski, {\it String
Theory}, Vol. 1, 2.}

\lref\papers{A.~Bergman, K.~Dasgupta, O.~J.~Ganor,
J.~L.~Karczmarek and G.~Rajesh, ``Nonlocal field theories and
their gravity duals,'' Phys.\ Rev.\ D {\bf 65} (2002) 066005,
hep-th/0103090; K.~Dasgupta and M.~M.~Sheikh-Jabbari,
``Noncommutative dipole field theories,'' JHEP {\bf 0202} (2002)
002, hep-th/0112064.}

\lref\GP{ E.~Goldstein and S.~Prokushkin, ``Geometric model for
complex non-K\"ahler manifolds with SU(3) structure,''
hep-th/0212307.}

\lref\toappear{ K.~Becker, M.~Becker, K.~Dasgupta, P.~S.~Green,
..., work in progress.}

\lref\MeessenQM{ P.~Meessen and T.~Ortin, ``An Sl(2,Z) multiplet
of nine-dimensional type II supergravity theories,'' Nucl.\
Phys.\ B {\bf 541} (1999) 195-245, hep-th/9806120.}

\lref\bem{P.~Bouwnegt, J.~Evslin and V.~Mathai, ``T-duality:
topology change and H flux'', hep-th/0306062.}

\lref\HUT{ C.~M.~Hull and P.~K.~Townsend, ``The two loop beta
function for sigma models with torsion,'' Phys.\ Lett.\ B {\bf
191} (1987)  115-121; ``World sheet supersymmetry and anomaly
cancellation in the heterotic string,'' Phys.\ Lett.\ B {\bf 178}
(1986) 187.}

\lref\SenJS{ A.~Sen, ``Dynamics of multiple Kaluza-Klein
monopoles in M and string theory,'' Adv.\ Theor.\ Math.\ Phys.\
{\bf 1} (1998) 115-126, hep-th/9707042. }

\lref\sav{ K.~Dasgupta, G.~Rajesh and S.~Sethi, ``M theory,
orientifolds and G-flux,'' JHEP {\bf 9908} (1999) 023,
hep-th/9908088.}

\lref\BeckerNN{ K.~Becker, M.~Becker, M.~Haack and J.~Louis,
``Supersymmetry breaking and $\alpha'$-corrections to flux
induced  potentials,'' JHEP {\bf 0206} (2002) 060,
hep-th/0204254.}

\lref\maeda{K.~i.~Maeda, ``Attractor in a superstring model: the
Einstein theory, the Friedmann universe and inflation,'' Phys.\
Rev.\ D {\bf 35} (1987) 471-479.}

\lref\rohmwit{R.~Rohm and E.~Witten, ``The antisymmetric tensor
field in superstring theory,'' Annals Phys.\  {\bf 170} (1986)
454-489.}

\lref\kg{S.~Gukov, S.~Kachru, X.~Liu, L.~McAllister, to appear.}

\lref\eva{E.~Silverstein, ``(A)dS backgrounds from asymmetric
orientifolds,'' hep-th/0106209; S.~Hellerman, J.~McGreevy and
B.~Williams, ``Geometric constructions of nongeometric string
theories,'' hep-th/0208174; A.~Dabholkar and C.~Hull, ``Duality
twists, orbifolds, and fluxes,'' hep-th/0210209.}

\lref\bbsb{K.~Becker and M.~Becker, ``Supersymmetry breaking,
M-theory and fluxes'' JHEP {\bf 0107} (2001) 038, hep-th/0107044.}

\lref\bbdg{K.~Becker, M.~Becker, K.~Dasgupta and P.~S.~Green,
``Compactifications of heterotic theory on non-K\"ahler complex
manifolds: I,'' hep-th/0301161.}

\lref\ibars{I.~Bars, D.~Nemeschansky and S.~Yankielowicz,
``Compactified superstrings and torsion,'' Nucl.\ Phys.\ B {\bf
278} (1986) 632-656; I.~Bars, D.~Nemeschansky and S.~Yankielowicz,
``Torsion in superstrings,'' SLAC-PUB-3775, presented at the {\it
Workshop on Unified String Theories}, Santa Barbara, Calif., Jul
29-Aug 16, 1985; I.~Bars, ``Compactification of superstrings and
torsion,'' Phys.\ Rev.\ D {\bf 33} (1986) 383-388.}

\lref\candle{P.~Candelas, G.~T.~Horowitz, A.~Strominger and
E.~Witten, ``Vacuum configurations for superstrings,'' Nucl.\
Phys.\ B {\bf 258} (1985) 46-74.}

\lref\lenny {L.~Susskind ``The anthropic landscape of string
theory,'' hep-th/0302219.}

\lref\mikei{M.~R.~Douglas, ``The statistics of string/M theory
vacua'' JHEP {\bf 0305} (2003) 046, hep-th/0303194;
S.~Ashok and M.~R.~Douglas,`` Counting flux vacua'',
hep-th/0307049.}

\lref\bdine{T.~Banks and M.~Dine, ``Is there a string theory
landscape'', hep-th/0309170.}

\lref\weba{J.~Bagger and J.~Wess, {\it ``Supersymmetry and
Supergravity,''} Princeton University Press.}

\lref\becons{ M.~Becker and D.~Constantin, ``A note on flux
induced superpotentials in string theory,'' hep-th/0210131.}

\lref\witsuper{E.~Witten, ``New issues in manifolds of $SU(3)$
holonomy,'' Nucl.\ Phys.\ B {\bf 268} (1986) 79-112.}

\lref\douglas{M.~R.~Douglas, ``The statistics of string/M theory
vacua,'' hep-th/0303194.}

\lref\senF{A.~Sen, ``Orientifold limit of F-theory vacua,''
Phys.\ Rev.\ D {\bf 55} (1997) 7345-7349, hep-th/9702165;
``Orientifold limit of F-theory vacua,'' Nucl.\ Phys.\ Proc.\
Suppl.\  {\bf 68} (1998) 92-98 [Nucl.\ Phys.\ Proc.\ Suppl.\
{\bf 67} (1998) 81-87], hep-th/9709159.}

\lref\gopmuk{R.~Gopakumar and S.~Mukhi, ``Orbifold and
orientifold compactifications of F-theory and M-theory  to six
and four dimensions,'' Nucl.\ Phys.\ B {\bf 479} (1996) 260-284,
hep-th/9607057.}

\lref\origa{S.~Chakravarty, K.~Dasgupta, O.~J.~Ganor and
G.~Rajesh, ``Pinned branes and new non Lorentz invariant
theories,'' Nucl.\ Phys.\ B {\bf 587} (2000) 228-248,
hep-th/0002175; K.~Dasgupta, G.~Rajesh, D.~Robbins and S.~Sethi,
``Time-dependent warping, fluxes, and NCYM,'' JHEP {\bf 0303}
(2003) 041, hep-th/0302049.}

\lref\toappear{ K.~Becker, M.~Becker, K.~Dasgupta, E.~Goldstein,
P.~S.~Green and S.~Prokushkin, work in progress.}

\lref\anna{A.~Fino and G.~Grantcharov, ``On some properties of
the manifolds with skew-symmetric torsion and holonomy SU(n) and
Sp(n),'' math.dg/0302358.}

\lref\bg{K.~Behrndt and S.~Gukov, ``Domain walls and
superpotentials from M theory on Calabi-Yau three-folds,'' Nucl.\
Phys.\ B {\bf 580} (2000) 225-242, hep-th/0001082.}

\lref\luest{G.~L.~Cardoso, G.~Curio, G.~Dall'Agata and D.~Lust,
``BPS action and superpotential for heterotic string
compactifications with fluxes,''
JHEP {\bf 0310}, 004 (2003),
hep-th/0306088.}

\lref\bbdp{K.~Becker, M.~Becker, K.~Dasgupta and S.~Prokushkin,
``Properties of heterotic vacua from superpotentials,''
Nucl.\ Phys.\ B {\bf 666}, 144 (2003),
hep-th/0304001.}

\lref\bbdg{K.~Becker, M.~Becker, K.~Dasgupta and P.~S.~Green,
``Compactifications of heterotic theory on non-K\"ahler complex
manifolds. I,'' JHEP {\bf 0304} (2003) 007, hep-th/0301161.}

\lref\disgreone{J.~Distler and B.~Greene, ``Aspects of (2,0)
string compactifications,'' Nucl. Phys. {\bf B304} (1988) 1-62.}

\lref\ct{C.~Callan and L.~Thorlacius, ``Sigma models and string
theory,'' in {\it Particles, Strings, and Supernovae}, the
proceedings of TASI 1988.}

\lref\gswtwo{M.~Green, J.~Schwarz, and E.~Witten, {\it Superstring
theory}, volume 2.}

\lref\mathaipriv{V. Mathai, private communication.}

\lref\liyau{J.~Li and S.-T.~Yau, ``Hermitian Yang-Mills
connections on non-K\"ahler manifolds,'' in {\it Mathematical
Aspects of String Theory}, World Scientific, 1987.}

\lref\tomaone{M.~Toma, ``Stable bundles on non-algebraic surfaces
giving rise to compact moduli spaces,'' C. R. Acad. Sci. Paris
S\'er. I Math. {\bf 323} (1996) 501-505.}

\lref\tomatwo{M.~Toma, ``Compact moduli spaces of stable sheaves
over non-algebraic surfaces,'' Doc. Math. {\bf 6} (2001) 11-29.}

\lref\kcsub{E.~Sharpe, ``K\"ahler cone substructure,'' Adv. Theor.
Math. Phys. {\bf 2} (1999) 1441-1462, hep-th/9810064.}

\lref\shamitjohn{S.~Kachru and J.~McGreevy, ``Supersymmetric
three-cycles and supersymmetry breaking,'' Phys. Rev. {\bf D61}
(2000) 026001, hep-th/9908135.}

\lref\ce{E.~Calabi and B.~Eckmann, ``A class of compact, complex
manifolds which are not algebraic,'' Ann. of Math. (2) {\bf 58}
(1953) 494-500.}

\lref\lutian{P.~Lu and G.~Tian, ``The complex structure on a
connected sum of $S^3 \times S^3$ with trivial canonical bundle,''
Math. Ann. {\bf 298} (1994) 761-764.}

\lref\reid{M.~Reid, ``The moduli space of 3-folds with $K=0$ may
nevertheless be irreducible,'' Math. Ann. {\bf 278} (1987)
329-334.}

\lref\wall{C.~T.~C.~Wall, ``Classification problems in
differential topology, V:  on certain 6-manifolds,'' Inv. Math.
{\bf 1} (1966) 355-374.}

\lref\blumZ{J.~D.~Blum and A.~Zaffaroni, {``An orientifold from F
theory,''} Phys.\ Lett.\ B {\bf 387}, 71 (1996), hep-th/9607019.}

\lref\fmw{R.~Friedman, J.~Morgan, and E.~Witten, ``Vector bundles
and F-theory,'' Comm. Math. Phys. {\bf 187} (1997) 679-743,
hep-th/9701162.}

\lref\dtthree{E.~Sharpe, ``Discrete torsion,'' hep-th/0008154.}

\lref\dtrev{E.~Sharpe, ``Recent developments in discrete
torsion,'' Phys. Lett. B {\bf 498} (2001) 104-110,
hep-th/0008191.}

\lref\dtshift{E.~Sharpe, ``Discrete torsion and shift orbifolds,''
Nucl. Phys. B {\bf 664} (2003) 21-44, hep-th/0302152.}

\lref\evaed{E.~Silverstein and E.~Witten, ``Criteria for conformal
invariance of $(0,2)$ models,'' Nucl. Phys. B {\bf 444} (1995)
161-190, hep-th/9503212.}

\lref\candelasetal{P.~Berglund, P.~Candelas, X.~de~la~Ossa,
E.~Derrick, J.~Distler, and T.~Hubsch, ``On the instanton
contributions to the masses and couplings of $E_6$ singlets,''
Nucl. Phys. B {\bf 454} (1995) 127-163, hep-th/9505164.}

\lref\chrised{C.~Beasley and E.~Witten, ``Residues and worldsheet
instantons,'' hep-th/0304115.}

\lref\bbh{K.~Becker, M.~Becker, M.~Haack, and J.~Louis,
``Supersymmetry breaking and $\alpha'$-corrections to flux
induced potentials,'' hep-th/0204254.}

\lref\ersrev{E.~Sharpe, ``Lectures on D-branes and sheaves,''
hep-th/0307245.}

\lref\tian{G.~Tian and S.~T.~Yau, ``Three-dimensional algebraic
manifolds with $c_1 = 0$ and $\chi = -6$.''}

\lref\candelas{P.~Candelas, P.~S.~Green and T.~Hubsch, ``Finite
distances between distinct Calabi-Yau Vacua: (Other worlds are
just around the corner),'' Phys.\ Rev.\ Lett.\  {\bf 62}, 1956
(1989); P.~Candelas, P.~S.~Green and T.~Hubsch, ``Rolling among
Calabi-Yau vacua,'' Nucl.\ Phys.\ B {\bf 330}, 49 (1990).}

\lref\andy{A.~Strominger, ``Massless black holes and conifolds in
string theory,'' Nucl.\ Phys.\ B {\bf 451}, 96 (1995),
hep-th/9504090.}

\lref\gms{B.~R.~Greene, D.~R.~Morrison and A.~Strominger,
``Black hole condensation and the unification of string vacua,''
Nucl.\ Phys.\ B {\bf 451}, 109 (1995), hep-th/9504145.}

\lref\mirror{B.~R.~Greene and M.~R.~Plesser, ``Duality in
Calabi-Yau moduli space,'' Nucl.\ Phys.\ B {\bf 338}, 15 (1990);
P.~Candelas, X.~C.~De La Ossa, P.~S.~Green and L.~Parkes, ``A pair
of Calabi-Yau manifolds as an exactly soluble superconformal
theory,'' Nucl.\ Phys.\ B {\bf 359}, 21 (1991); P.~Candelas,
X.~C.~De la Ossa, P.~S.~Green and L.~Parkes, ``An exactly soluble
superconformal theory from a mirror pair of Calabi-Yau
manifolds,'' Phys.\ Lett.\ B {\bf 258}, 118 (1991).}

\lref\kstt{S.~Kachru, M.~B.~Schulz, P.~K.~Tripathy and
S.~P.~Trivedi, ``New supersymmetric string compactifications,''
JHEP {\bf 0303}, 061 (2003), hep-th/0211182.}

\lref\gaugL{G.~L.~Cardoso, G.~Curio, G.~Dall'Agata and D.~Lust,
``Heterotic string theory on non-K\"ahler manifolds with H-Flux
and gaugino condensate,'' hep-th/0310021.}

\lref\Kgukov{S.~Gukov, S.~Kachru, X.~Liu and L.~McAllister,
``Heterotic moduli stabilization with fractional Chern-Simons
invariants,'' hep-th/0310159.}

\Title{\vbox{\hbox{hep-th/0312221} \hbox{UMD-PP-04-15}
\hbox{SU-ITP-03/34}}} {\vbox{ \vskip-2.5in
\hbox{\centerline{K\"ahler versus Non-K\"ahler
Compactifications}}}}

\vskip-.2in
\centerline{\bf Melanie Becker${}^1$, ~~~Keshav Dasgupta${}^2$}
\vskip.2in
\centerline{\it ${}^1$~Department of Physics, University of
Maryland, College Park, MD 20742-4111} \vskip.02in
\centerline{\it ${}^2$~Department of
Physics, Varian Lab., Stanford University, Stanford CA 94305-4060}
\vskip.02in
\centerline{\tt melanieb@physics.umd.edu, keshav@itp.stanford.edu}

\vskip.6in

\centerline{\bf Abstract}

\noindent We review${}^\ast$ our present understanding of
heterotic compactifications on non-K\"ahler complex manifolds
with torsion. Most of these manifolds can be obtained by duality
chasing a consistent F-theory compactification in the presence of
fluxes. We show that the duality map generically leads to
non-K\"ahler spaces on the heterotic side, although under some
special conditions we recover K\"ahler compactifications. The
dynamics of the heterotic theory is governed by a new
superpotential and minimizing this superpotential reproduces all
the torsional constraints. This superpotential also fixes most of
the moduli, including the radial modulus. We discuss some new
connections between K\"ahler and non-K\"ahler compactifications,
including some phenomenological aspects of the latter
compactifications.

\vskip3cm

\noindent
$\underline{~~~~~~~~~~~~~~~~~~~~~~~~~~~~~~~~~~~~~~~~~~~~~~~~~~~~~}$
\Date{\it ${}^\ast$ Based on the talks given at the QTS3
conference, University of Cincinnatti and SUSY 03.}

The Calabi-Yau (CY) compactifications of Candelas et al. \candle\
have led to some major progress in our understanding of string
theory vacua. Compactifying the heterotic string on such manifolds
results in four-dimensional models with minimal supersymmetry
(susy). In terms of the corresponding two dimensional non-linear
sigma model, we demand conformal invariance so that all the
tadpoles vanish and the string equations of motion are satisfied.
In this way we recover again CY spaces. For the bosonic case this
will give us the model studied in \nemes. By definition CY
manifolds are K\"ahler and have a vanishing first Chern class. By
Yau's theorem therefore, for a {\it given} complex structure and
a {\it given} cohomology class of the K\"ahler form there is a
{\it unique} Ricci-flat metric with $SU(3)$ holonomy\foot{Ricci
flatness is not an essential property of the compactifying manifold as
has been demonstrated in \senn. We can restore K\"ahlerity
without having a Ricci flat metric (a field redefination relates them).
For the non-K\"ahler
manifolds, however, we can never have a Ricci flat metric.}.

Generically, when considering ordinary CY compactifications of the
heterotic string theory the three-form background fluxes (at weak 
coupling and constant dilaton) are equal to zero. The four-dimensional spacetime is
Minkowski and therefore has zero cosmological constant. Although
susy would also allow
 anti de-Sitter solutions, here only the
Minkowski solution is realized. The cancellation of the two-loop
sigma model beta function puts a strong constraint on the vector
bundle, namely it has to be identified with the tangent bundle,
implying that the three form is in the cohomology classes of the
manifold \foot{This is a sufficient condition, but not a necessary
one. We shall discuss this in more detail as we go along.}. Further, by the Uhlenbeck-Yau theorem, there is an
essentially unique choice of vector field for any {\it given}
holomorphic stable vector bundle satisfying the
Donaldson-Uhlenbeck-Yau (DUY) equations.

This attractive scenario however is clouded by some inherent
problems which are related to the degeneracy of string vacua.
Essentially there are two different degeneracies appearing in
string theory compactifications. First, there are thousands of CY
manifolds that could be potential solutions to the low energy
effective theory. Second, once we choose a {\it particular} CY
manifold, there are many different moduli associated with the
complex structure and the K\"ahler structure deformations of the
manifold. All these moduli are unstable at tree level and thus
lead to a situation that is unattractive for phenomenology. It
turns out that the radial modulus of the CY is one of the
K\"ahler moduli. Therefore, when this field is not stabilized the
CY will runaway to infinite size. This ruins the whole
consistency of the compactification scenario \DineSB.

One way to remove the degeneracy for a given CY is to relax the
restriction on the fundamental two-form $J$ by allowing manifolds
that have $dJ \ne 0$, i.e spaces that are non-K\"ahler. These more
general compactifications were first discussed in detail in
\rstrom,\HULL\ and \SmitD. 
At first sight it may not be apparent at all how one could remove degeneracies
by going to non-K\"ahler manifolds. This will be explained below. But first observe that
breaking the K\"ahler condition is not
straightforward, as one can show that in the absence of
background fluxes {\it and} warped metrics, the generic solution
is always K\"ahler. Therefore for a non-K\"ahler manifold to be a
solution of the equations of motion one has to switch on non
trivial three form fluxes ${\cal H}$, which will essentially play the role of
a torsion. The torsion is not closed because an embedding of the
spin-connection into the gauge-connection is not allowed, as this
would lead back to CY manifolds. Furthermore, the dilaton is
generically non constant and related to the warp factor of the
underlying manifold.

The first concrete example of such manifolds was constructed in
\sav\ and \beckerD\ by duality chasing a particular model of the
general class of M-theory compactifications with non vanishing
fluxes considered in \rBB. The manifold constructed is a $T^2$
bundle over a four dimensional $K3$ base. In \beckerD\ and \bbdg\
many properties of this manifold were explicitly found by going to
the orbifold limit of the $K3$ base. So for example, the manifold
is compact, complex, has a vanishing first Chern class and $SU(3)$
holonomy. It was observed in \GP, \bbdg\ that the Betti numbers
of this manifold are different from the Betti numbers of a simple
product $K3 \times T^2$ (appearing for vanishing fluxes), implying
a topology change. The topology change is achieved by considering
an additional {\it twist} (along with the flux) so that a
consistent solution that preserves minimal susy in four
dimensions is obtained. Due to this fact it is {\it not} possible
to construct this manifold directly in the heterotic theory using
a supergravity analysis. This has been explained in \bbdp. However
the duality chasing that we performed miraculously takes the
topology change into account, so that a consistent non-K\"ahler
manifold appears on the heterotic side. See \beckerD\ and \bbdp\
for a more detailed analysis of this. In fact, the {\it twist}
that we expect on the heterotic side is actually one component of
the $G$-flux in M-theory.

An immediate advantage that compactifications on manifolds with
torsion have is moduli stabilization at tree level. The fluxes
give rise to a potential that stabilizes {\it all} the complex
structure moduli \beckerD, the radial modulus and some of the
remaining K\"ahler moduli \bbdp, \bbdg. This stabilization can be
understood in terms of a superpotential first constructed in
\bbdp\ and verified later by dimensional reduction in \luest.
Contrary to popular belief, the superpotential is complex and is
given by $W = \int ({\cal H} + i dJ) \wedge \Omega$. Here $J$ is
the fundamental two form which may not necessarily be integrable
for an arbitrary choice of fluxes and $\Omega$ is the unique
holomorphic (3,0) form wrt the almost complex structure
that characterizes these manifolds. The
form of this superpotential implies that the dynamics of the
heterotic theory compactified on these non-K\"ahler manifolds can
be described by a {\it complex} three form ${\cal G}$. This three
form is anomaly free and gauge invariant (see \bbdg\ and \bbdp\
for a derivation of this) and therefore can be used to construct
the scalar potential for all the moduli from its kinetic term
$\int \vert {\cal G} \vert^2$. This potential incorporates terms
that are of higher order in $\alpha'$ (see \bbdgs\ for details).
Notice, that the no-scale structure of the potential is broken in
a rather interesting way. The anomaly free three form can be shown
to depend secretly on the radial modulus (and some of the K\"ahler
moduli) by solving the anomaly condition \bbdp. The non-trivial
radial dependence comes from the fact that the Bianchi identity
incorporates the three form on both sides of the equation and
therefore can be solved iteratively order by order in $\alpha'$
\bbdg, \bbdp. In \bbdp\ a simple analysis was performed to
evaluate the radius of the non-K\"ahler manifold. It was shown
that the $K3$ base can be made large enough by choosing large
flux densities, but the fiber generically has a size of order
$\alpha'$. This again implies that a simple supergravity analysis
cannot be performed directly in the heterotic theory and duality
chasing needs to be performed \foot{Although the duality chasing
works for most cases, sometimes it may not give the complete
answer. For example, the full heterotic Bianchi identity, or the
scalar potential do not follow from simple duality chasing. The
derivation of these require higher order $\alpha'$ corrections of
the action and the T-duality rules, that do not appear in the
supergravity approximation. These subtleties have been explained
in \bbdg\ and \bbdgs.}.

There are a couple of questions one might ask at this point. One
immediate one would be whether we can choose {\it any} fourfold in
M-theory, or whether this choice is restricted. As in the usual
relation between compactifications of F-theory on a fourfold and
the heterotic string on a threefold there is, of course, the
restriction that the fourfold should be an elliptic fibration.
Observe that the particular fourfold that was chosen for duality
chasing i.e $K3 \times K3$, has an orientifold description on the
type IIB side, which makes the description easier. The orientifold
operation actually involves three actions: world sheet parity
(the usual orientifold), fermion number reversal and space
reversal (the orbifold action). The presence of these three
actions guarantees that a set of U-dualities will take us to the
heterotic string \sav,\beckerD\ and not bring us back to the
strongly coupled type IIB theory. This aspect was used in \bbdgs\
to construct new examples of non-K\"ahler manifolds that have non
zero Euler characteristics. The original non-K\"ahler manifolds
constructed in \sav,\beckerD\ have zero Euler characteristics
\GP, \bbdg.

The next question is the choice of fluxes. Anomaly cancellation in
M-theory implies that fluxes are necessary, if compactifications
on manifolds with non-vanishing Euler characteristics are being
considered \rBB, \SethiVW. An alternative picture in which fluxes
are traded with space filling branes and their consequent effect
on the geometry is discussed in \bbdgs. From this discussion one
could naively conclude that one should {\it always} get a
torsional compactification on the heterotic side and the simpler
CY compactifications are ruled out. Interestingly, it turns out
that for some special choice of fluxes, we will get back an
ordinary CY compactification. Let us elaborate this a little bit.
More details will appear elsewhere. There are two different types
of fluxes that we can choose on the M-theory side: one that is
positioned over the full fourfold and the other that is {\it
localized} at the fixed points of the fourfold. These fixed
points are the points where the $T^2$ fibration degenerates. As
discussed in detail in \sav, \beckerD, the non-localized fluxes
appear on the type IIB side as the NS and RR three form fields.
If these fluxes are not vanishing we eventually obtain the
non-K\"ahler manifolds with torsion three form on the heterotic
side after further U-dualities. The localized fluxes, on the other
hand, become the seven brane gauge fields on the type IIB side
\beckerD, \bbdg, which under further U-dualities become the
heterotic gauge fields. These gauge fields originate in M-theory
by decomposing the G-fluxes in terms of the localized
(normalizable) harmonic (1,1) forms near each singularity
\robbins. In the presence of non-localized G-fluxes the localized
(1,1) forms themselves change by the backreaction of the fluxes on
the geometry. This can be worked out with some effort \robbins,
but will be ignored in the following. Generically in the presence
of both fluxes we would get non-K\"ahler spaces with torsion.
Furthermore, the background G-fluxes {\it warp} the geometry in
some special way \sav,\beckerD,\bbdg. The equation for the warp
factor was given in \sav,\beckerD. What happens if we choose only
the localized fluxes? It is easy to see that in type IIB theory we
will get gauge fields on the seven branes and no three form
background fluxes. The anomaly cancellation condition will put
some restriction on the total instanton numbers of these gauge
fields. This is of course the usual restriction that we expect
also on the heterotic side. Therefore we seem to recover ordinary
CY compactifications, except for the fact that in the presence of
fluxes in M-theory we will typically get a warped metric. This
would naively ruin the $dJ = 0$ property. However a careful
analysis reveals that the warp factor equation e.g. on the type
IIB side, is in fact proportional to $\Omega_3(\omega) -
\Omega_3(A)$, where $\Omega_3$ is the Chern-Simons three form (see
\sav\ and \beckerD\ for the derivation of this). Therefore if we
{\it embed} the spin connection $\omega$ into the gauge connection
$A$ we will recover a trivial warp factor and the manifold will
become K\"ahler! This is precisely the reason behind embedding the
spin-connection into the gauge connection. Here we have rederived
this property from M-theory by demanding the consistency of CY
compactifications.

There is another rather interesting aspect that comes to mind at
this point. Imagine that we do not turn on the non-localized gauge
fluxes and at the same time do not allow the standard embedding.
Then the naive expectation would be that we should get a
non-K\"ahler manifold with the non-K\"ahlerity coming precisely
from the difference $\Omega_3(\omega) - \Omega_3(A)$. This would
seem to contradict the result of \Kgukov\ where it was found that
a fractional gauge Chern-Simons term can appear in an ordinary CY
compactification. However this apparent puzzle can be resolved by
taking the background {\it gaugino condensate} into 
account\foot{The gaugino condensate contributes to the (3,0) and the 
(0,3) part of the threeform, and therefore breaks susy.}. To
see how this helps we need to back up a little for more generality.

The complex three form that appears in the heterotic theory in the
presence of torsion has to be imaginary self-dual (ISD) to
preserve supersymmetry in four dimensions. This implies the
background equation $dJ = \ast {\cal H}$, where $\ast$ is the Hodge
duality in six dimensions (over the non-K\"ahler manifold). This
equation is more general than the constraint derived in
\rstrom,\HULL, \SmitD\ and it reduces to the known form when the
manifold is complex (see \bbdp\ for a derivation of this fact).
This equation makes the non-K\"ahler nature of the manifold
manifest. Observe that if we scale the metric then the three form
${\cal H}$ scales linearly too. On the other hand from the Bianchi
identity we observe, that the three form does not scale (at least
to the lowest order in $\alpha'$). This implies that the radial
modulus should get stabilized, as we saw earlier. This argument,
although correct, is rather naive at this point. The fact that
the Bianchi identity does not scale is only true for the K\"ahler
case. In the non-K\"ahler case since the three form appears on
both sides of the identity, the issue is more subtle. Therefore
the correct way to study the potential for the radial modulus
would be to evaluate the three form flux order by order in
$\alpha'$ 
and use the kinetic term to calculate the potential.
This was done in \bbdp, \bbdgs.

Coming back to the torsional constraint, we see that in the
absence of three form fluxes the manifold can still become
non-K\"ahler via the relation $dJ = \ast~\alpha'[\Omega_3(\omega)
- \Omega_3(A)]$. In \Kgukov\ the $\Omega_3(\omega)$ term was
cancelled by one of the Chern-Simons terms of the gauge fields. It
turns out that one can make $dJ = 0$ using the gaugino condensate
contribution to the superpotential. The gaugino condensate will
change the torsional equation by an additional term (for a
derivation of this see for example \gaugL). This additional term
can be used to cancel the remaining gauge Chern-Simons part and we
recover the K\"ahler property\foot{A more ``dynamical'' way to
achieve this would be to take our proposed superpotential and
solve for $dJ$ using the various contributions (tree level,
perturbative and non-perturbative). If $dJ = 0$ with integrable
$J$ we get K\"ahler CY compactifications. All other cases, i.e
when $dJ \ne 0$ with $J$ integrable or non-integrable, will
correspond to non-K\"ahler compactifications. As we saw earlier,
for this case the radius is stabilized at tree level. For the
K\"ahler case, the radius is fixed non-perturbatively
\buchkov,\Kgukov. In both cases the $\sigma$-model conformal
invariance is restored at {\it this} particular radius.}.

Therefore we see that K\"ahler compactifications are a special
case of the more general non-K\"ahler compactifications with
torsion (at least those non-K\"ahler compactifications that could
be constructed from F-theory using duality chasing). To summarize:
the most generic superpotential governing these backgrounds is
complex and only for the special case $dJ = 0$ we recover the
superpotential proposed in \witsuper.

Our next goal is to understand the fibration structure of our
non-K\"ahler manifold. The precise metric of the fiber torus has
been worked out in \beckerD, \bbdg. As we discussed earlier,
there is a change in Betti numbers when we go from $K3 \times T^2$
to the non-K\"ahler space. It turns out that this change of
``topology'' can be understood from a {\it brane-box}
configuration in the type IIB theory. The brane-box divides the
region in two parts: one that is inside the box and the other
outside. The walls of the box are made out of NS5-branes which are
actually sources of NS flux. The RR three form flux can be
obtained from the T-dual twist of a type IIA configuration (see
\bbdgs\ for a detailed analysis of this scenario)\foot{As an
additional advantage we get non constant three form fluxes, as
opposed to the constant fluxes of \sav,\beckerD.}. Under a set of
U-dualities each side of the brane box transforms into a Taub-NUT
space which reproduces the fibration structure. The fact that the
metric of the system works out correctly has been checked in
\bbdgs. To see how the superpotential works out from the brane box
configuration we observe that the RR three form sources can also
be replaced by D5 branes (forming, say, the two other sides of
the box). These D5 branes become, under a set of U-dualities, NS5
branes wrapping some two cycles of the non-K\"ahler space whose
fibration structure is determined from the U-dual brane box. The
NS5 branes show a jump of the ${\cal H}$ charge precisely as ${\cal H} = - \ast
dJ$ \gauntlett, and therefore contribute the complex $dJ$ part of
the superpotential. More details on this will appear in a future
publication. Thus we can reproduce the full structure of the
non-K\"ahler space using a brane configuration. This should come
as no
 surprise because far away
 from the brane box configuration there is no distinction from the
 geometrical picture and the brane picture.
There are some subtleties in this construction primarily related
to the hidden orientifold nature of the system. The box
configuration survives this projection. But one also has to take
into account the F-theory monodromies to reproduce the complete
gauge bundles. These monodromies appear as a stringy cosmic string
in our model (again for more details the readers are advised to
look in \bbdgs).

Before moving ahead, we still have to clarify the reason of why
these
 manifolds have an $SU(3)$ holonomy (wrt the torsional connection).
The existence of minimal susy in four dimensions gives us a
necessary condition that is, however, not sufficient. A
slightly more
 stronger argument\foot{We thank Paul Green for providing the
following  argument.}
 will be to observe that if we choose the metric so that the cube
 of $J$ is the product of the
holomorphic three form $\Omega$ and its dual, then the holomorphic
connection that respects the metric will also respect the
holomorphic three form and therefore will have  SU(3)
 holonomy. The
profound aspect of Yau's theorem is that, in the CY case, the
metric can
 be chosen to be
K\"ahler in addition to the above property. Since there is no
K\"ahler metric for the more general non-K\"ahler
compactifications, we have nothing comparable to prove. We do,
however, need another principle for choosing the particular metric
we have chosen out of an infinite dimensional space of metrics.
The above criterium is satisfied by our choice of metric.

We can also explain this using the torsion classes ${\cal W}_i$
(with $i = 1,... , 5$) of \carluest\ and \louisL. In this
classification, a non-K\"ahler manifold that preserves an $SU(3)$
holonomy will have to necessarily satisfy the equation $2{\cal
W}_4 + {\cal W}_5 = 0$, with an additional condition ${\cal W}_1
= {\cal W}_2 = 0$, so that the complex structure is integrable.
For more details on the physics aspects of this see for example
\carluest. For our non-K\"ahler manifold the torsion classes have
been worked out in \bbdg\ and for the Iwasawa manifold, in
\carluest. Using this analysis, it can also be shown that one {\it
cannot} define any K\"ahler metric on these manifolds (see
\GP,\bbdg\ for more details). Therefore these manifolds are
explicitly non-K\"ahler.

So far our discussion has been mostly restricted to non-K\"ahler
manifolds that have zero Euler characteristics. Having zero Euler
characteristics is not much of a problem, of course, because we
are not embedding the spin connection into the gauge connection.
The reason why we would like to look for more general manifolds
with non-zero Euler characteristics is purely to extend our
understanding of these manifolds. This may serve as a new and
interesting direction in mathematics and may also turn out to be
phenomenologically more attractive.

The first interesting example of such a manifold is four
dimensional and can be described in terms of a $K3$ manifold with
torsion. The metric has been worked out in \rstrom, where it was
shown to pick up an overall conformal factor from the back
reaction of the fluxes on the geometry. One may wonder if it is
possible to obtain this manifold from duality chasing in
F-theory. Indeed this has been achieved in \bbdgs\ (at least for
the non-compact case). Therein it was shown that the metric is
related to a Gimon-Polchinski kind of model with the
axion-dilaton $\tau$ fixed at a {\it particular} value. In this
sense this is different from our earlier examples where we had a
vanishing axion-dilaton and therefore F-theory was at {\it
constant} coupling \dasmukhi. Now since the axion-dilaton is non-trivial,
there would be sizable non-perturbative corrections to the model.
These corrections actually convert the intersecting orientifold
planes and branes to smooth hyperbolas. For details on this see
\sengimon\ for the case of zero torsion and \bbdgs\ for the case
with torsion. An alternative way to appreciate this would be to
observe that the metric of a torsional $K3$ is precisely the
metric of a NS5 brane at a point on the K3. Therefore under a set
of U-dualities the system maps to a configuration of $T^4$
orientifolds with seven branes. These seven branes are the
sources of non trivial $\tau$ in this framework (see \bbdgs\ for
more details).

There are many subtleties that we have ignored here, which have
been discussed in \bbdgs\ though. The generic orientifold action
in the type IIB case is in fact {\it ambiguous}. This ambiguity
gives different heterotic duals and is present in four- as well as
six-dimensional models. Of course, once we fix the heterotic
compactification, we also fix the type IIB ambiguity.
Nevertheless, the six-dimensional compactification becomes rather
involved because of this subtlety. For a more detailed discussion
of this issue in the generic context see \gopmuk, and for the
torsional case, see \bbdgs.

This brings us to the next interesting case of a six-dimensional
compactification with non zero Euler characteristics. The duality
chains from F-theory to the heterotic theory were given in \bbdgs.
If we keep the orientifold action $\Gamma$ as generic, then the
type IIB manifold will be of the form ${\cal N}_6/\Gamma$, where
${\cal N}_6$ is a six-dimensional compact manifold (in the absence
of fluxes) that has non zero Euler characteristics. In the
presence of fluxes, the heterotic dual of this manifold will look
like a $Z_2$ action of a torus fibration (locally) over a compact
base. If we choose another action of $\Gamma$ that gives a $P^1$
fibration over a $P^1 \times P^1$ base on the type IIB side (in
the absense of fluxes), then the heterotic manifold will be a
non-trivial six manifold that is a $Z_2$ action of a torus
fibration over a $P^1 \times P^1$ base. The generic fibration
structure is not difficult to work out (though the analysis may
get very tedious). The result, in a compact form, is presented in
\bbdgs\ and therefore we refer the reader to this paper. However,
the full topological data have not yet been worked out. This will
be presented elsewhere.

The examples discussed above are all of orbifold nature but in
principle smooth examples do exist. Some of these have been
discussed in \bbdgs. They include the connected sums of $S^3
\times S^3$ (first discussed in \papado)
and some examples of flops of an elliptically fibered
CY space. The flops used herein are the ones that break the
K\"ahler condition. In \bbdgs\ it was discussed in detail how the
bundles follow the manifold through the flop. But whether these
smooth examples are solutions of string equations of motion have
not been discussed yet. This is relegated to a future publication.

The question of finding stable vector bundles for our manifolds is
a very important one especially because we are no longer allowed
to embed the spin connection into the gauge connection (see
\PapaDI, \bbdg\ for a discussion on this). The DUY equations take
the same form as in the CY case, i.e the (2,0) and the (0,2) part
of the curvature vanishes and the (1,1) part is traceless.
However, there are two additional conditions. First, notice that
$J$ is no longer closed (and may not be integrable either) and
second, there is now a constraint on tr($F \wedge F$) coming from
the $i \del \bar\del J$ part (see \bbdgs\ for details). These
conditions look very restrictive and one might wonder if there
exists {\it any} solution at all to these equations. Again the
duality chasing comes to the rescue here. The Weierstrass equation
governing the F-theory background allows an $D_4^4$ bundle to
propagate to the heterotic side taking the orientifold action
into account. This bundle does satisfy all the conditions as has
been explicitly demonstrated in \bbdg. One has to carefully take
the orbifold singularities, localized fluxes {\it and} higher
order corrections in F-theory into account. Failing to do so will
not reproduce the correct result \bbdg. In this way we obtain
one particular consistent example, for which the above equations
can be solved.

There are several open questions that might be raised at this
point. First is the question of the {\it stability} of the
bundle\foot{Earlier work on this appeared in \liyau.}. For
K\"ahler compactifications the holomorphic gauge fields that
satisfy the DUY equations are equivalent to Mumford-Takemoto
stable holomorphic bundles. For non-K\"ahler compactifications,
assuming the metric to be (approximately) Gauduchon, a similar
statement can also be made (though under some special
circumstances). For more details see \bbdgs. Unfortunately a full
understanding of the bundles has not yet been achieved and we
hope to address this question in a near future.

Another open question is to understand the full non-abelian nature
of the bundle. From the duality chasing one expects the full
non-abelian nature to show up. But in practice, it is only the
abelian part that is manifest in this scenario (the localized
G-fluxes form the Cartan subalgebra of the $D_4$ algebra). The
full non-abelian part can be seen, if we consider the M2 branes
to be wrapping the degenerating cycles of Taub-NUT. In \bbdg\ the
intersection matrices of these two cycles were shown to reproduce
exactly the Cartan matrix of the $D_4$ algebra (at least near one
of the four orbifold singularties).

Finally, one might ask about the number of generations for these
non-K\"ahler models. Since we do not allow the standard embedding,
the number of generations is {\it not} equal to the Euler number
of the non-K\"ahler manifold. However we need to evaluate the
number of generations when (a) the non-trivial warp factor is
taken into consideration, and (b) the spin connection is the
torsional connection (in addition to not having standard
embedding). Somewhat surprisingly, the number of generations is
still given by the third Chern class $c_3$ of the bundle. For a
proof of this see \bbdgs. In \bbdgs\ the number of generations
for a rather simple example with a $U(1)$ bundle has been worked
out. For a more realistic non-K\"ahler compactification one would
have to compute the number of generations for a phenomenologically
relevant group (say for example $SU(5)$). This would mean that we
need an $SU(5)$ bundle on our non-K\"ahler space satisfying the
DUY equations. Whether this is indeed possible remains to be
seen. More details on this will be reported elsewhere.

Having a detailed analysis of non-K\"ahler manifolds, it is now
time to ask whether we can formulate a non-linear sigma model
description of these manifolds. Some details of this have
appeared in the early works of \rstrom,\HULL, where the basic
constraints were shown to follow from a (0,2) sigma model
description (for a more recent exposure see \bbdg\ and \bbdgs).
One would expect a reasonable non-linear sigma model description
behind our construction but the calculation of massless spectra is
rather subtle here. See \bbdgs\ for a discussion on this and the
issue of ${\cal H}$-twisted sheaf cohomology for the counting of states.
However at this point, it is not clear how to formulate a simple
{\it linear} sigma model that flows to the conformally invariant
background that we have.

The non-linear sigma model, on the other hand, can be efficiently
used to describe various interesting aspects of the torsional
backgrounds. In particular one can explicitly derive the preferred
connection for this background. The preferred connection is chosen
in such a way that with this choice the heterotic $\sigma$-model
action resembles the Green-Schwarz superstring action in a
background configuration with 24 free fields. The identification
becomes precise as soon as we embed the gauge connection in the
modified spin-connection. However, as we discussed above, the
standard embedding is not allowed here; and therefore only to the
lowest order in $\alpha'$ the two actions may be identified. On
the other hand, with this choice of connection one can in fact use
the Green-Schwarz superstring action to compute ${\cal
O}(\alpha')$ corrections to the heterotic beta functions. All
these corrections are suppressed by the size of the six-manifold,
and  conformal invariance is restored when the manifold attains
the size that is dictated by the minima of our superpotential.
For this and other details regarding $\sigma$-model description
see \HULL\ for an early exposition and \bbdg\ for a more recent
discussion related to the superpotential. The fact that three form
appears on both sides of the Bianchi identity can also be easily
shown. The torsional equation and the DUY relations follow from
demanding world sheet susy \HULL,\rstrom, \bbdg. One can also
show that the fundamental form should be ${\cal H}$-covariantly
constant, so that the manifold allows an integrable complex
structure.

The above discussion, hopefully, summarizes our present
understanding of heterotic compactifications on non-K\"ahler
complex manifolds with torsion. We have shown that it is possible
to get an almost {\it rigid} vacua by using background fluxes. To
finish this summary let us remark that there is another important
issue that has not been addressed so far. This has
to do with the {\it number} of possible string vacua. Recall that
at the beginning of this note we discussed two kind of
degeneracies: one of them originated from the fact that many
different manifolds can be solutions to the string equations of
motion and the other one resulted from the deformations of a {\it
given} vacuum. What we discussed so far concerns only the lifting
of the second type of degeneracy i.e. the fixing of the moduli
for a {\it given} vacuum. But we have, at present, no
understanding on how to fix the other degeneracy.
A recent counting of flux vacua in a different context has
revealed that this number could be finite in some cases \mikei.
It would be rather interesting to understand if such a
calculation can be performed in the present scenario.

\centerline{\bf Acknowledgements}

\noindent It is our pleasure to thank K.~Becker, E.~Goldstein,
P.~S.~Green, S.~Prokushkin and E.~Sharpe for their collaboration
leading to some of the results presented herein. Furthermore we would
like to thank the organizers of QTS3 and SUSY 03 for organizing a
stimulating conference. The work of M.B. is supported by NSF
grant PHY-01-5-23911 and an Alfred Sloan Fellowship. The work of
K.D is supported by a David and Lucile Packard Foundation
Fellowship 2000-13856.


\listrefs

\bye